\documentstyle[12pt]{article}
\def\TL{\hfil$\displaystyle{##}$}
\def\TR{$\displaystyle{{}##}$\hfil}

\def\comment#1{}
\def\fixit#1{}

\def\mop#1{\mathop{\rm #1}\nolimits}

\def\coth{\mop{coth}}

\def\sech{\mop{sech}}

\def\lbldef#1#2{\expandafter\gdef\csname #1\endcsname {#2}}
\def\eqn#1#2{\lbldef{#1}{(\ref{#1})}%
\begin{equation} #2 \label{#1} \end{equation}}
\def\eqalign#1{\vcenter{\openup1\jot
    \halign{\strut\span\TL & \span\TR\cr #1 \cr
   }}}

\newcommand{\newsection}[1]{
\vspace{10mm}
\pagebreak[3]
\addtocounter{section}{1}
\setcounter{subsection}{0}
\noindent
{\large\bf \thesection. #1}
\nopagebreak
\medskip
\nopagebreak}
\newcommand{\newsubsection}[1]{
\vspace{5mm}
\pagebreak[3]
\addtocounter{subsection}{1}
\addcontentsline{toc}{subsection}{\protect
\numberline{\arabic{section}.\arabic{subsection}}{#1}}
\noindent{\em 
\thesubsection. #1}
\nopagebreak
\vspace{2mm}
\nopagebreak}
\def\comment#1{  \begin{raggedright}{\tt [#1]}\end{raggedright}}
\def\fixit#1{}

\def\comment#1{  \begin{raggedright}{\tt [#1]}\end{raggedright}}
\def\fixit#1{}

\def\half{{ 1 \over 2 } }
\def\tt{\tilde{t}}
\def\l{\ell}
\def\vzs{\vec{z}^{\, 2}}

\def\({\left(}
\def\){\right)}
\def\[{\left[}
\def\]{\right]}
\def\tt{\tilde{t}}
\def\tH{\tilde{H}}
\def\p{\partial}
\def\R{{\bf R}}
\def\S{{\bf S}}
\def\CN{{\cal N}}
\def\CO{{\cal O}}
\begin{document}
\baselineskip=15.5pt
\pagestyle{plain}
\setcounter{page}{1}


\begin{titlepage}

\begin{flushright}
PUPT- 2031 \\  
SU-ITP-02/21 \\
hep-th/0205258
\end{flushright} 
\vfil

\begin{center}
{\Large Penrose Limits and Non-local theories}\\
\end{center}
\vspace{1.5cm}
\begin{center}
{\large Veronika E. Hubeny$^a$, Mukund Rangamani$^b$ and Erik Verlinde$^b$}
\end{center}
\begin{center}   
$^a$ Physics Department, Stanford University, Stanford, CA 94305. USA \\
$^b$ Department of Physics, Princeton University, Princeton,
NJ 08544. USA 
\end{center}  
\vfil

\begin{center}
{\large Abstract}
\end{center}
\noindent
We investigate Penrose limits of two classes of non-local theories,
little string theories and non-commutative gauge theories. Penrose limits 
of the near-horizon geometry of NS5-branes help to shed some 
light on the high energy spectrum of little string theories.  
We attempt to understand renormalization group flow in these 
theories by considering Penrose limits wherein the null geodesic also 
has a radial component. In particular, we demonstrate that it is 
possible to construct a pp-wave spacetime which interpolates between 
the linear dilaton and the AdS regions for the Type IIA NS5-brane.
Similar analysis is considered for the holographic dual geometry to 
non-commutative field theories.

\vfil
\vfil
\vspace{6cm}
\begin{flushleft}
May 2002.
\end{flushleft}

\end{titlepage}
\newpage 

\newsection{Introduction}

String theory has provided natural candidates for non-gravitational theories 
which are nonetheless non-local. Prominent examples in this 
category of theories are non-commutative field/string theories
\cite{cds, dh, sw,sst,gmms} and little string theories \cite{brs, seiberg}. 
Understanding the details of dynamical behaviour in 
these theories might shed light on the 
consequences of non-local interactions in a quantum theory,
thereby hopefully paving the way toward  quantum gravity.

A very useful tool to understand the dynamics of these non-local theories is  
to use their holographic description, wherein one can talk about string 
propagation in a background spacetime. In the case of little string 
theories one is led to considering string theories in linear dilaton 
backgrounds \cite{abks}. For non-commutative theories,
the holographic dual is the near horizon geometry of the D-brane 
configuration realizing the gauge theory in the low energy limit.
In the case of 4-dimensional non-commutative theories, this is a
specific deformation of the $AdS_5$ background \cite{hait,maru}.
The holographic description has been useful
to understand some of the characteristic features of these theories. For 
instance, one can compute correlation functions of operators 
in the supergravity limit, as in 
\cite{abks, minsei, NM} for little string theories and as in \cite{maru} for 
non-commutative theories.

While the holographic description has many positive features, a full  
solution of string theory in the backgrounds in question is lacking. So it
is fair to say that one doesn't have a complete picture of the 
dual theory. However, it might be possible that if one isolated a certain 
sector of the field theory, one could hope for a solution within 
that particular sector. To this end, one needs to identify a solvable sector 
of the field theory, which {\it a priori} is not an easy task. 
Instead, one can draw inspiration from the gravity dual.

In general relativity, just as one can consider zooming 
in onto a point in a manifold, thus giving rise to the important concept of 
tangent space, one can consider zooming in onto a null geodesic in a spacetime.
While the former is flat, the latter has more structure, essentially
coming from the direction along the geodesic; in particular, it gives rise
to a curved space known as a plane-parallel wave, or pp-wave, 
as was demonstrated by Penrose \cite{penrose} several decades ago.

More specifically, in taking the so-called Penrose limit of a spacetime, 
one boosts to the speed of light, which would typically tend 
to make the curvature tensor components diverge, but at the same time, one 
also scales up the metric by ``zooming in'' infinitesimally close to 
the geodesic, which scales the curvature back down.  
One may also regard this as the limiting procedure in which one considers
timelike observers going faster and faster, but with their clocks appropriately
recalibrated. By scaling up a coordinate 
transverse to the direction along the geodesic, one obtains a Killing field;
in fact, one can show that this will be a covariantly 
constant null Killing field. This is the essential 
ingredient to the pp-wave geometry.

Recently, a very interesting solvable model of string theory in 
Ramond-Ramond backgrounds was proposed in \cite{metsaev,Bmn}. 
This was achieved by taking the Penrose limit \cite{penrose} of 
$AdS_5 \times \S^5$ spacetime \cite{Bmn,tseytlin}, the 
holographic dual of $ d=4$, $\CN =4$ Super-Yang-Mills theory. In this limit 
one finds that one is concentrating on a particular sector of the gauge theory,
with large dimensions of operators and large charges, but with a finite 
difference between the charges and the dimensions. Since the dual background 
is exactly solvable as a string theory, we can claim to have understood, 
at least in principle, this particular sector of the gauge theory. 

Since the new scaling limit of ${\cal N} = 4$ Yang-Mills defined in \cite{Bmn}
yielded new insights into the gauge/string correspondence, 
we can ask whether a similar 
strategy will lead to something useful in the context of little strings. 
In fact, the answer is affirmative and we will shortly see how we can 
obtain some information about the high energy spectrum of little strings. 
This will be achieved by considering the Penrose limit of the 
linear dilaton geometry, such that on the little string side we are 
concentrating on the sector of the theory with energies of order 
$\sqrt{N} l_s$ and carrying a $U(1)$ charge of order $N$. 
$N$ is the usual dimensionless number characterizing the 
little string theory, which we shall take large so as to trust the 
supergravity approximation.
For this sector of the little string theory, we will find that 
the spectrum is just the free string spectrum in 10 dimensions.

We then turn to an investigation of renormalization group trajectories 
in the non-local theories in  question. Under the holographic map, 
the renormalization group flow in the field theory side is associated 
with the radial evolution in the spacetime geometry. Thus, if we were to 
consider null geodesics which probe the radial direction, 
we could gain some intuition about the RG flows in the dual field theory.  
We will construct Penrose limits of null geodesics with radial component in 
the cigar geometry to study the RG flow in little string theory with an 
infra-red cut-off. In addition, we will show that in the pp-wave limit, it is 
possible to construct an RG trajectory that interpolates between the 
linear dilaton region and the $AdS_7$ region corresponding to the 
infra-red $(2,0)$ superconformal field theory, for the case of 
the Type IIA little string theory. We will see that the resulting pp-wave 
geometries are time-dependent, this time dependence being interpreted as  
the scale dependence of the dual theory. A similar situation can 
be studied for non-commutative theories and we also comment on 
constructing such geometries.

The organization of the paper is as follows. We begin in Section 2
with a general introduction to Penrose limits, concentrating 
on a particular class of metrics which we  use in the later part of the 
paper. In Section 3, we  discuss the Penrose limit of the linear 
dilaton geometry and show that one can extract the high energy spectrum of the 
little string theory from the resulting plane wave geometry. Section 4 and 5 
 deal with understanding the basic properties of renormalization group
flow in little string theories. In Section 4, we  consider a null geodesic 
in the cigar geometry, which is dual to the little string theory at finite 
energy densities, while in Section 5, we  consider a more general solution 
stemming from an array of M5-branes in M-theory. Section 6  deals with 
similar exercise for non-commutative theories, concentrating in particular 
on the case of the D3-brane. Finally, we  end with some conclusions.

\newsection{Penrose Limits}

As mentioned in the introduction, the Penrose limit is defined as the 
spacetime in the neighbourhood of a null geodesic in a specific scaling limit,
which we will discuss in detail below. The limit was first proposed in 
\cite{penrose} in the context of classical general relativity 
and was extended to supergravity solutions in \cite{gueven}. 
Recent discussions including many examples of Penrose limits for 
D-brane geometries can be found in \cite{blaua,blaub,blauc}. We shall in the 
next subsection review the general formulation of the Penrose limit, 
following \cite{gueven}.

\newsubsection{The Penrose limit}

Given any spacetime with a metric and a null geodesic in the 
spacetime, one can define the Penrose limit in the following fashion.
We first choose our coordinate system so that it is well adapted to the 
null geodesic. This may be done by choosing a coordinate $u$ which is 
the affine parameter along a congruence of  null geodesics,  
a coordinate $v$ to be the distance between neighbouring geodesics,  
and finally $x^i$ to denote the remainder of the coordinates. 
These coordinates are chosen so as to 
satisfy $g_{uu} =0, g_{u x^i }= 0$ and $g_{uv} =  N $
so that the general form for the metric is 
\eqn{uvform}{
ds^2 = N \left(2 \, du \, dv + dv \left( dv + \sum_i \, B_i(u,v,x^i)  
dx^i \right)+  
\sum_{i,j} \, C_{ij}(u, v, x^i) \, dx^i \, dx^j \right),
}
where $N$ is a dimensionless parameter introduced to keep track of the scaling
limit of various quantities. Later on, when we discuss spacetime backgrounds
of D-branes, $N$ will be the number of D-branes, which we will choose large 
so as to have a well behaved semi-classical approximation.
In the Penrose scaling limit \cite{penrose}
we define the following rescaling of coordinates:
\eqn{uvrescale}{
\left(u, v,x^i \right) \rightarrow \; \left(u, {v \over N},{x^i \over \sqrt{N}}
\right),
}
and take $N \rightarrow \infty$. In this limit we see that the $dv^2$ and
the $dv \, dx^i$ terms drop out and we are left with the pp-wave metric
\eqn{rosenpp}{
ds^2 =  2 \, du \, dv +  \sum_{i,j} \,C_{ij}(u) \, dx^i \, dx^j
}
Note that the metric functions $C_{ij}$ now become functions 
of the coordinate $u$ alone. This form of the metric is called the 
Rosen form \cite{rosen}.

However, this metric is typically not geodesically complete: one can't 
cover the full spacetime without encountering coordinate singularities.
As these are not physical singularities, however, one can write the
metric in a more suitable form which does cover the full spacetime, 
as noticed by \cite{robinson}.  This is called the Brinkman \cite{brinkman},
or harmonic, form, and is given by
\eqn{Brinkman}{
ds^2 = 2 \, du \, dv + \( \sum_{i,j} A_{ij}(u) \, x^i \, x^j \) du^2
+ \sum_{i} dx^i \, dx^i}
This form has the virtue of exhibiting the deviations from flat spacetime
more clearly, and it turns out to be more amenable to doing light-cone
quantization in string theory.

It is worthwhile to note that there is an equivalent method to define the 
pp-wave limit. Specifically, instead of adapting the coordinate $u$ to a 
congruence of null geodesics, one simply considers it to be the affine 
parameter along a single geodesic, so that $g_{uu} =0$ only along the geodesic,
and $ g_{u x^i }= 0$, $g_{uv} =  N $ as before.
In the scaling 
limit, $dv^2$ and the $dv \, dx^i$ terms drop out, but we are left with
a $du^2$ term, as well as some non-flat $dx^i \, dx^i$ terms. The resulting 
pp-wave metric will partially be in Rosen and partially in Brinkman form.
We will find this procedure calculationally more convenient in the 
rest of the discussion.

One can explicitly write down the coordinate transformations 
relating the Brinkman and Rosen forms for the metric, 
as done {\it e.g.}\ in \cite{blauc}. We will present the relevant formulae
for a simple metric ansatz with the metric functions $C_{ij}(u)$ 
appearing in \rosenpp\ being diagonal. Let us write the metric in Rosen 
coordinates as 
\eqn{nonBr}{
ds^2 = 2 \, d{\bar u} \, d{\bar v} + G({\bar u}) \, d{\bar x}^2.} 
Then the following change of variables 
\eqn{ctBr}{
{\bar u} = u, \ \ 
{\bar x} = {x \over \sqrt{G(u)}}, \ \ {\rm and } \ \ 
{\bar v} = v + {1\over 4} {\p_u G(u) \over G(u)} \, x^2 ,
}
casts the metric in the form 
\eqn{Br}{
 ds^2 =  2 \, du \, dv + dx^2 + {1\over 2} \, x^2 
\( {1\over 2} {\(\p_u G(u)\) ^2 \over G(u)^2} + \p_u \( {\p_u G(u) \over G(u)}
 \) \) \, du^2.}
This gives the pp-wave metric in the desired Brinkman form.
 
\newsubsection{Penrose limit for a metric ansatz}

Let us now see how the Penrose scaling limit works for a general spacetime
of the type we will consider in the remainder of the paper. For illustrative 
purposes, we will keep the notation simple while considering a general 
enough form of the metric, specifically
\eqn{genmetric}{
 ds^2 = N \[ -f(r) \, dt^2 + h(r) \, dr^2 + g(r) 
\( \cos^2 \theta \, d\psi^2 + d\theta^2 + \sin^2 \theta \, d\Omega_q^2  \) \] 
+ dy_{p}^2}
where $f$, $h$, and $g$ are arbitrary smooth functions of $r$.
This is a static $(p+q+4)$-dimensional metric, which is spherically symmetric
in $q+4$ directions and flat in remaining $p$ directions, expressed in a 
diagonal form. Typically, we will consider the metric as arising from a 
stack of $N$ D-branes; we will ultimately be interested in the large-$N$ 
limit. For further notational simplicity, we'll now drop the subscripts $p$ 
and $q$.

Consider a null geodesic in the  $(t,r,\psi)$ plane, at a fixed 
point in the $y$ space and around the equator of the 
transverse $q$-sphere, {\it i.e.},\ $y^i \equiv 0, \ \theta=0$.  
That such a geodesic is admissible, is obvious from the spacetime symmetries 
$y^i \to - y^i$ and $\theta \to -\theta$.
This geodesic is generated by the tangent vector 
$ \, \dot{t} \, \p_t + \dot{r} \, \p_r +\dot{\psi} \,\p_{\psi} $, 
the dot denoting derivative with respect to the affine parameter 
$\lambda$ along the geodesic, $\dot{} \equiv {d \over d \lambda}$.

Since $\p_t$ and $\p_\psi$ are Killing vectors as apparent from \genmetric,
 when contracted with a tangent vector to a geodesic, 
they define constants of motion along that geodesic.
In particular, $ E \equiv  f(r) \, \dot{t} $ and $J \equiv g(r) \cos^2 \theta \,
\dot{\psi}$ are conserved quantities, corresponding 
to the energy and angular momentum. For a null geodesic, $p^2 = 0$, 
which yields the following equation for $\dot{r}$ at $\theta = 0$:
\eqn{gengeod}{
\dot{r}^2 - {1 \over h(r)} \[ {E^2 \over f(r)} - {J^2 \over g(r)} \] = 0}
The second term defines an effective potential $V(r)$ 
for a one-dimensional motion of a particle, which one can easily analyze.
In particular, the only geodesic at constant $r= r_0$ is such that 
$V'(r_0) = 0$.

{\it A priori}, \gengeod\ describes a 2-parameter family of geodesics, 
determined by the constants of motion $E$ and $J$.
However, we can rescale these by redefining the affine parameter 
so that we have only one physical parameter, $\l \equiv J/E$, describing 
the geodesic. This is always possible for null geodesics because 
the scale drops out. Furthermore, from \gengeod\ we also see that $\l$ 
is bounded: for $h(r) > 0$, $\l^2 \le {g(r) \over f(r)}$ for all 
reachable $r$. We will assume that $\l > 0$, and return to the case 
of $\l = 0$, {\it i.e.}, purely radial geodesics, at the end of this section.

Let us now rewrite the metric in more suitable coordinates, adapted to 
the null geodesic. The coordinate transformation from
$(t,r,\psi) \to (u,v,x)$ is achieved by\footnote{ Here we are choosing 
{\em outgoing} null geodesics, corresponding to $\dot{r} > 0$;
however, we could have just as well picked the incoming ones.
Later, when we study a black hole geometry, this corresponds
to geodesics emerging from the white hole, which is the equivalent of
the geodesics falling into the black hole.}:
\eqn{genctd}{\eqalign{
dr &= {1 \over \sqrt{h(r)}} \sqrt{{1 \over f(r)} - {\l^2 \over g(r)}} \, du \cr
dt &= {1 \over f} \, du - dv + \l \, dx \cr
d\psi &= {\l \over g} \, du + dx
}}
We can use the first equation to solve for
\eqn{genur}{
u(r) = \int { \sqrt{h(r)} \over \sqrt{{1 \over f(r)} - {\l^2 \over g(r)}} } 
\, dr}
and invert to obtain $r = r(u)$.
The remaining functions $f$ and $g$ appearing in the second two equations of 
\genctd\ are then to be viewed as functions of $u$.

Substituting this back into the metric \genmetric, we obtain
\eqn{gensub}{\eqalign{
ds^2 = N \, & \Bigg( - f
\[ {1 \over f} \, du - dv + \l \, dx \]^2 
+ \( {1 \over f} - {\l^2 \over g} \) du^2 + \cr 
& \hspace{2cm} + g \left\{ \cos^2 \theta \, \[ {\l \over g} \, du + dx \]^2 
+ d\theta^2 + \sin^2 \theta \, d\Omega^2 \right\}  \Bigg) + dy^2 \cr
= N \, & \Bigg[ 2 \, du \, dv -  {\l^2 \over g} \sin^2 \theta \, du^2 
+ \( - \l^2 f + g - g \, \sin^2 \theta \) dx^2 
- 2 \l \, \sin^2 \theta \, du \, dx + \cr
& \hspace{ 2cm} + \dots dv^2 + \dots  dv \, dx  
+ g \( d\theta^2 + \sin^2 \theta \, d\Omega^2 \) \Bigg] + dy^2
}}
where the $\dots$ in the above will drop out momentarily.
We now rescale the coordinates as suggested by \uvrescale :
\eqn{coordres}{
u \to u, \;\;\;\; 
v \to {v \over N}, \;\;\;\; 
\theta \to {z \over \sqrt{N}}, \;\;\;\; 
x \to {x \over \sqrt{N}}, \;\;\;\; 
\Omega \to \Omega, \;\;\;\; 
y \to y}
Finally, taking the Penrose limit $N \to \infty$, 
yields the pp-wave metric
\eqn{largeNmet}{
ds^2 = 2 \, du \, dv - {\l^2 \over g} \, z^2 \, du^2 
+ \( - \l^2 f + g \) dx^2 + g \, \( dz^2 + z^2 \, d\Omega^2 \) + dy^2}
To simplify notation further, we rewrite the metric in the space transverse
to the equator on the original $q$-sphere as
$dz^2 + z^2 \, d\Omega^2 = d\vzs$,
so that $z^2 = \vzs$.  The pp-wave metric then takes the simple form
\eqn{largeNmetz}{
ds^2 = 2 \, du \, dv - {\l^2 \over g} \, \vzs \, du^2 
+ \( - \l^2 f + g \) dx^2 + g \, d\vzs + dy^2}

The metric as written contains functions $f(r(u))$ and $g(r(u))$ implicitly 
as functions of the coordinate $u$. The explicit form of these may be quite
complicated, and may not even have an analytic expression.
Also note that, written in this implicit form, the above expression
is independent of the function $h$. However, since the relation between 
$r$ and $u$ as determined by \genur\ depends on $h$, this is illusory.

To cast the metric in the Brinkman form we can follow the procedure 
outlined in the previous subsection. In particular, by applying the 
coordinate transformation suggested by \nonBr\ and \ctBr\ iteratively, 
the final metric \largeNmetz\ becomes
\eqn{genpp}{
ds^2 = 2 \, du \, dv 
+ \( {1 \over 2} F_z \, \vzs + {1 \over 2} F_x \, x^2 \) du^2 
+ dx^2 + d\vzs + dy^2}
with
\eqn{Fzx}{\eqalign{
F_z &= -2 {\l^2 \over g} 
      + \( {g' \over g} \)' + {1 \over 2} \( {g' \over g} \)^2 \cr
F_x &=  \( {(g-\l^2 f)' \over g-\l^2 f} \)' 
      + {1 \over 2} \( {(g-\l^2 f)' \over g-\l^2 f} \)^2 
}}
where the prime denotes differentiation with respect to $u$, 
$' \equiv {d \over du}$.

Now let us briefly return to the special case of radial geodesic, $\l = 0$.
The geodesic equation \gengeod\ now simplifies to 
$\dot{r}^2 = {1 \over f(r) \, h(r)}$, and correspondingly the 
coordinate transformation $(t,r) \to (u,v)$ is given by
\eqn{genradctd}{\eqalign{
 dr &= {du \over \sqrt{f(r) \, h(r)} } \cr
dt &= {1 \over f} \, du - dv}}
The metric expressed in these coordinates is then simply
\eqn{genradsub}{
ds^2 = N \[ 2 \, du \, dv - f \, dv^2 
+ g \( \cos^2 \theta \, d\psi^2 + d\theta^2 + \sin^2 \theta \, 
d\Omega^2  \) \] + dy^2}
Rescaling the coordinates as in \coordres\ and taking the large $N$ limit, 
we obtain the metric,
\eqn{genradpp}{
ds^2 = 2 \, du \, dv + g(u) \( dx^2 + d\vzs \) + dy^2}
Note that this is automatically expressed in the Rosen form.

\newsection{Ultra High Energies and Little Strings}

Little string theory (LST) is one of the most important examples of non-local
theories. This theory arises on the world-volume of 
NS5-branes when one considers a decoupling limit, $g_s \rightarrow 0$ with 
fixed $\alpha'$ \cite{brs, seiberg}. LSTs share many 
properties with usual string theory, such as T-duality and Hagedorn 
behaviour of density of states, but are nevertheless non-gravitational 
theories. As such, one wishes to uncover as many of their secrets as possible.
Presently, we will see that we can use the techniques outlined above 
to learn something about the high energy spectrum of the theory.

To probe the high energy regime we will consider the holographic dual 
of little string theory, formulated in terms of string 
propagation in the linear dilaton background \cite{abks}. 
We will zoom into the high energy sector of the little string 
theory by considering a particular Penrose limit of the spacetime.
Let us start by recalling some facts about the near-horizon 
geometry of NS5-branes.

The full NS5-brane metric in string frame and the dilaton are given by
\eqn{nsmetric}{\eqalign{
ds_{str}^2 & = -dt^2 + dy_5^2 + A(r) (dr^2 + r^2 d\Omega_3^2) \cr
e^{2\Phi} & = g_s^2 A(r)
}}
with
\eqn{harmonic}{
 A(r) = 1 + {N l_s^2 \over r^2}
}
In addition the supergravity solution also has a NS-NS three form $H$, 
which is $N$ times the volume form of the three sphere, where 
$N$ is the number of NS5-branes.

The near-horizon limit ($r \to 0$) of the NS5-brane metric 
\nsmetric, is achieved by dropping the $1$ in the harmonic function.
This is the linear dilaton geometry,
\eqn{nsfnhrest}{
ds_{str}^2 = N\, l_s^2\, \(\;
 -d\tilde{t}^2 + \cos^2 \theta \, d\psi^2 + d\theta^2 + 
\sin^2 \theta \, d\phi^2 + {dr^2 \over r^2} \;\)  + dy_5^2 
}
where we have rescaled the time coordinate,
$t = \sqrt{N} l_s \, \tilde{t} $, and written out the metric on the 
$\S^3$  as 
$d\Omega_3^2 = d\theta^2 + \cos^2\theta d\psi^2 +\sin^2\theta d\phi^2$, with 
$\theta =0$ being the equator.
Let us henceforth set $l_s \equiv 1$ and measure everything in the 
string units.
Roughly speaking, the radial direction, which is the linear dilaton direction,
holographically corresponds to the scale of the LST.

\newsubsection{Penrose limit of linear dilaton geometry}

To consider the appropriate pp-wave limit, 
we want to zero in onto a null geodesic in the above geometry \nsfnhrest.
Since we first want to consider the LST at a fixed scale,
we want our geodesic to sit at a constant value of the radial coordinate.
Our null geodesic will therefore
run along the equator of the three sphere transverse 
to the NS5-branes. This is the exact analog of the limit considered in 
\cite{Bmn} for $AdS_5 \times \S^5$, the near-horizon geometry of D3-branes.
Using \gengeod, we can check that there does exist 
a geodesic at constant value of the radial coordinate, provided the 
the angular momentum is saturated, namely $ \l =1 $.

We now introduce new coordinates $(u,v,x,z)$, making no changes to the 
coordinates $\phi$ and $y_5$, 
\eqn{lccoods}{\eqalign{
\tilde{t} \pm \psi & = 2 u,2 {v \over N} \cr
r &=  \sqrt{N} \; e^{x/\sqrt{N}} \cr
\theta &= {z \over \sqrt{N}} 
}}
where, as explained above,
we have chosen the coordinate $u$ to be  the affine 
parameter along the null geodesic, and the coordinate $v$ 
to define a  null, covariantly constant Killing vector in the plane wave limit.
After taking the large $N$ limit, keeping the 
rescaled coordinates fixed,
we obtain the metric
\eqn{lcmetric}{
ds_{str}^2 = -4 \, du \, dv - \vzs du^2 + d\vzs + dx^2 + dy_5^2
}
where $d\vzs = dz^2 + z^2 \, d\phi^2$. We note in passing that 
a trivial rescaling of $v$ 
will cast the metric in \lcmetric\
into the standard form of the pp-wave metric given in \genpp.

Also, in this limit, the dilaton becomes a constant, 
\eqn{lcdilaton}{
e^{2\Phi} = g_s^2 e^{x/\sqrt{N}} \rightarrow g_s^2, 
}
and the non-trivial 
three-form NS-NS field scales to $H_{u 12} = 3$, where by indices
$(1,2)$ we denote the coordinates in the $\vec{z}$ plane, which we will
henceforth parametrize as $(z^1, z^2)$. The fact that we have a constant 
$H$-field implies that $B_{12} = u$ on the world-sheet. 

Finally, with a trivial 
rescaling, we can write the Penrose 
limit of the near-horizon geometry of the NS5-branes as
\eqn{penroselim}{\eqalign{
ds_{str}^2 &= -4 \, du \, dv -\mu^2 \vzs \, du^2 + d\vzs + dx^2 + dy_5^2 \cr
B_{12} &=  \mu\, u 
}}
This defines an exact string background to all orders in 
world-sheet perturbation theory \cite{amkl, horst} and was 
first derived by \cite{nappiw} and recently rederived in the 
context of Penrose limits of NS5-brane backgrounds by \cite{googuri}
and in \cite{kp}.

\newsubsection{Implications for Little String Theory}

Thus far, we have considered a particular scaling of the metric
corresponding to the near-horizon geometry of NS5-branes. To make 
contact with little string theory, we need to identify the precise meaning 
of the Penrose limit from the dual theory point of view. 

We can relate the light-cone energy and momentum to the 
original energies and momenta as measured in the
{\it  string frame metric} \nsmetric.
Recalling that $\p_t = {1 \over \sqrt{N}} \p_{\tilde t}$,
\eqn{lcenergies}{\eqalign{
2p^{-} &= -i {\partial \over \partial u} = -i\({\partial \over \partial 
\tilde{t}} + {\partial \over \partial \psi} \) = \tilde{E} - J =
 \sqrt{N} E - J \cr
2p^{+} &= -i {\partial \over \partial v} = -{i \over N}
\({\partial \over \partial 
\tilde{t}} -{\partial \over \partial \psi} \) = {\tilde{E} + J \over N}=
{ \sqrt{N} E + J \over N}
}}
If, in the original metric \nsmetric, we look at energies $E$ of order 
$\sqrt{N}$ and angular momenta of order $N$, with the additional 
constraint that $\sqrt{N} E - J $ is finite, then we can describe all 
such states in little string theory by string propagation in 
the Nappi-Witten background \penroselim. Thus, the claim is that 
the spectrum of strings in the background \penroselim\ is isomorphic 
to the little string spectrum in the regime $\sqrt{N} E \sim J \sim \CO(N)$.

It is very simple to read off the spectrum of string theory in 
the background \penroselim, 
\cite{kk} - \cite{russotb}.
The presence of the covariantly constant,  
null Killing vector allows us to quantize the theory in light-cone gauge
\cite{horowitzs}. We identify the world-sheet time $\tau$ with the 
light-cone coordinate $u$, $ u = \tau$. In light-cone gauge, the 
Polyakov sigma model reduces to six free 
bosons and two bosons $(z^1,z^2)$ which are ``massive'' and have a time 
dependent magnetic field. Let us concentrate on the latter, for the 
former fields are quantized in the usual fashion.

The light-cone Hamiltonian for the two bosons $(z^1, z^2)$ is 
\eqn{lcham}{\eqalign{
H_{lc} = {1 \over 4 \pi \alpha'} \,
\int d\tau \; \int_0^{2\pi\alpha' p^+} d\sigma \;
&\left(\half \left(\partial_\tau z^1\right)^2 - 
\half \left(\partial_\sigma z^1\right)^2 - \half \mu^2 (z^1)^2
+ z^1 \leftrightarrow z^2 \right) +\cr
& \hspace{2.5cm} i \mu \,\left(\tau \partial_\tau z^1 \partial_\sigma z^2 -
  z^1 \leftrightarrow z^2 \right)
}}
By a change of variables these can be recast 
as free bosons with shifted oscillators. To see this, note that we 
can do away with the magnetic field term and the mass term by 
writing the lagrangian in terms of $Z = e^{i \mu \sigma} (z^1 + i z^2)$.
The complex scalar field $Z$ is then a free scalar on the world-sheet.

The shift is a result of the time dependent magnetic field on the 
world-sheet. The spectrum then is very similar to 
free string theory, modulo the shift, since at the end of the day we are 
left with free scalars.

The upshot is that the spectrum of little string theory in the 
very high energy regime is almost free string spectrum in 
flat space in ten dimensions. We are here considering energies of 
order $\sqrt{N} l_s$ in the sector with $U(1)$ charge $J \sim N$. 
This was a hitherto unexplored regime of little string theory, and 
our result yields a prediction that little string theory ought to have states  
whose properties are identical to states of free string theory in 
10 dimensions.

\newsection{Penrose limits in the  cigar}

One of the problems with string propagation in the linear 
dilaton background is that as we move further down the throat, we 
encounter a local string coupling that is diverging, rendering the 
string perturbation analysis invalid. We actually got around the same in the 
previous section, by a judicious choice of null geodesic. By staying put at 
constant radial coordinate in the near-horizon metric \nsfnhrest, we 
managed to stay away from the strong coupling region. This involves 
choosing the constant radial coordinate to be large -- precisely what we 
need if we are going to talk about the high energy spectrum of little 
string theory.

We now would also like to incorporate some effects of the non-trivial 
geometry in the radial direction arising due to the presence of the dilaton. 
However, to start with, we want to excise the strong coupling region. Later 
we shall also incorporate the strong coupling effects by studying the 
appropriate M-theory background, but for now, we want to stay within the 
realm of ten-dimensional supergravity for simplicity of the discussion. 
Such a geometry is immediately available to us in the form of  
two-dimensional black hole. This is the near-horizon limit of a non-extremal
NS5-brane taken in an appropriate scaling limit. 
 
The near-horizon
geometry of the non-extremal NS5-brane is given by the metric
\eqn{nonextns}{\eqalign{
 ds_{str}^2 &= -\left(1- {r_0^2 \over r^2}\right) \, dt^2 + dy_5^2  + 
 { N l_s^2 \over r^2}
 \left({dr^2 \over 1 - {r_0^2 \over r^2} } +
 r^2 d\Omega_3^2 \right) \cr
e^{2\Phi} &= g_s^2 \, { N l_s^2 \over r^2} 
}}
The decoupling limit \cite{malstr} is defined as a double-scaling limit, 
such that the asymptotic value of the string coupling and the horizon 
radius are scaled to zero simultaneously, 
while keeping fixed the energy density above 
extremality in string units, {\it i.e.}, 
\eqn{decolim}{
g_s \to 0, \;\;\;\; r_0 \to 0, 
\;\;\;\;\; \mu_{lst} = {r_0^2 \over g_s^2 \ l_s^2} = \rm{fixed}
}
To analyze this limit, it is convenient to 
rescale  the time coordinate $t= \tilde{t} \, \sqrt{N} l_s $ and
introduce a new radial coordinate
$r = r_0 \cosh \sigma$, which in the scaling limit gives 
the near-horizon geometry
\eqn{scalim}{\eqalign{
ds_{str}^2 &=  N l_s^2 \left( -\tanh^2 \sigma \;d\tilde{t}^2 + 
d\sigma^2 + 
d\Omega_3^2  \right)+ dy_5^2 \cr
 e^{2\Phi} &= {N \over \mu_{lst} \ \cosh^2 \sigma} 
}}
Again, we can measure all quantities in the string units, 
so that $l_s \equiv 1$. The coordinate $\sigma$ is adapted to 
the region outside the black hole horizon.

The geometry described by \scalim\ is a convenient way to regulate the 
strong coupling region. One interpretation of this geometry is to work in the 
Euclidean signature black hole, and then this is the holographic dual of 
little string theory at finite temperature; the temperature being held 
fixed at the Hagedorn value of $T_H = {1 \over 2 \pi \sqrt{N \alpha'}}$.
We could also work with the Lorentzian cigar, which 
corresponds to little string theory at finite energy density.

\newsubsection{Null geodesics, scalings and Penrose limit}

We will now examine the Penrose limit of this geometry.
First of all, let us write the geodesic equation.
As before, we parameterize the 3-sphere by
 $d\Omega_3^2 = d\theta^2 + \cos^2\theta \, d\psi^2 + \sin^2\theta \,
d\phi^2$ and consider a geodesic in the $(\tt, \sigma, \psi)$ plane,
along the equator $\theta = 0$ of the 3-sphere.
Proceeding in the usual way explained in Section 2, we arrive at the 
geodesic equation,
\eqn{resgeodesic}{
-{1 \over \tanh^2\sigma } + \({d\sigma \over du}\)^2 + \l^2 = 0 
}

A few points can be noted from this equation. If we insist on 
a geodesic at a constant value of $\sigma$, {\it i.e.}, look for a 
massless particle orbiting around the transverse three sphere without 
radially moving in toward (or away from) the black hole, then the only such 
location is at $\sigma = \infty$. This corresponds to the asymptotically 
linear dilaton part of the geometry, where the presence of the black hole is 
irrelevant. This will lead to the geometry which we have studied in the 
previous section, the Nappi-Witten background.

Our aim will be to explore the modifications caused by the presence 
of the black hole. One interesting avenue which suggests itself
for exploration is to look at null geodesics with a radial component 
in order to understand what happens to the little string theory under 
RG flow, since according to the usual holographic map,
the radial motion ought to correspond to scale transformations. 
If we consider purely radial 
null geodesics, {\it i.e.}, geodesics with no angular momentum on the 
three sphere, then using \resgeodesic\ and the results of Section 2, it is 
easy to see that we end up with flat space with a linear dilaton proportional 
to $u$. This is an uninteresting geometry, as it encodes 
little information about the little string theory. Therefore we will 
consider geodesics which are 
radially infalling, as well as carrying some angular momentum along the
3-sphere. As we will see, the corresponding Penrose limit 
will lead to pp-wave geometries with a time dependent metric.

To proceed, we solve for the affine parameter as a function of the 
radial coordinate; a straightforward integration yields 
\eqn{sigmau}{
\cosh(\sigma) =  
 {\l \over \sqrt{1-\l^2}} \; \sinh \( \sqrt{1-\l^2} \, u \) }
Note that this implies that $\cosh(\sigma) =  u$ in the limit $\l = 1$.
The limiting case $\l \to 1$ is clearly a special case, since then the
geodesic equation simplifies dramatically. 
Physically, this corresponds to the angular momentum being saturated.

Let us therefore first derive the explicit form of the Penrose limit metric
for this special case, $\l =1$. The change of coordinates suggested by
\genctd\ is
\eqn{ctrans}{\eqalign{
\sigma &= \cosh^{-1} u \cr
\psi &= u + x \cr
\tt &= (u - \coth^{-1} u ) -v + x
}}
Substituting the above coordinate change and scaling the coordinates as in
\coordres\ followed by the large $N$ limit gives:
\eqn{Penrc}{
ds^2 = 2 \, du \, dv - \vzs \, du^2 + {dx^2 \over u^2} 
+ d\vzs  + dy_5^2}
where we have used $\vec{z}$ to encode the $(z,\phi)$ plane, as in \largeNmetz.
For later convenience, we recast this into the Brinkman coordinates 
{\it via} the  coordinate transformation \ctBr,
\eqn{ctcig}{
x \to ux, \ \ \ \  v \to v - {x^2 \over 2u},}
so that the resulting metric becomes
\eqn{Penrf}{
ds^2 = 2 \, du \, dv + \( {2x^2 \over u^2} - \vzs \) du^2 + dx^2
+ d\vzs + dy_5^2.
}

Now let us go back and consider the more general case $0 \le \l \le 1$.
Let $\cosh(\sigma) ={ \l \over \sqrt{1-\l^2}} \; \sinh \(\sqrt{1-\l^2}\, u \) 
 \equiv g(u)$.
The appropriate change of coordinates which generalizes \ctrans\ is given by 
integrating
\eqn{ctransl}{\eqalign{
d\sigma &= \sqrt{{g(u)^2 \over g(u)^2 -1} - \l^2} \  du \cr
d\psi &= \l \, du + dx \cr
d\tt &= {g(u)^2 \over g(u)^2 -1}\ du -dv + \l \, dx
}}
Now, we go through the same procedure as above and obtain at the 
end of the day, the pp-wave metric in Brinkman coordinates as:
\eqn{Penrfl}{
ds^2 = 2 \, du \, dv 
+ \(2 { (1-\l^2) \over \sinh^2(\sqrt{1-\l^2}\, u ) }\, x^2 
- \l^2 \, \vzs \) du^2 + dx^2
+ d\vzs + dy_5^2,}
Note that all of the above formulae reduce to the previous ones in the limit
$\l \to 1$. The $\l = 0$ case is a singular limit; 
to keep the original $\cosh(\sigma)$ finite, 
$\sinh(\sqrt{1-\l^2} \, u) 
\to \infty$.  In this limit, the whole $du^2$ term vanishes, in agreement
with our generic result \genradpp.

The backgrounds \Penrf\ and \Penrfl\ also have a non-trivial dilaton 
which is a function of the $u$ coordinate alone, and a 
NS-NS $B$-field given as ,
\eqn{dilBcigar}{\eqalign{
e^{2 \Phi(u)} & = {N \over \mu_{lst}} \, {1- \l^2 \over \l^2 
\sinh^2(\sqrt{1-\l^2} \, u)} \cr
B_{z_1 z_2 } & = \l \, u 
}}

One can understand the geometry \Penrfl\ as follows: in the Penrose limit
we are zooming into the part of the spacetime in the neighbourhood of
a massless particle trajectory. The coordinate $u$ parametrizes the 
trajectory, and is basically the radial coordinate in the original 
cigar metric \scalim. The $u$ dependence of the metric \Penrfl\ is 
explained easily, for by construction we are considering 
geodesics moving along that direction.
By virtue of zooming in on our geodesic around the equator
of the $\S^3$, we have effectively decompactified it, and the $\R^2$
parametrized by $\vec{z} = (z^1, z^2)$. 
The longitudinal directions of the five-brane $y_5$ play no interesting 
role and simply go along for the ride. We note that the only difference 
of this metric from the Nappi-Witten background derived in the previous 
section is the term proportional to $x^2$ in $g_{uu}$. Importantly, in the 
limit $u \rightarrow \infty$ this vanishes, as it must from our
discussion following the analysis of the geodesic equation \resgeodesic.
We note that similar time-dependent pp-wave backgrounds were derived in 
\cite{zayas}.

\newsubsection{String Propagation}

To study string propagation in \Penrf\ and  \Penrfl\ we will write the 
sigma model in light-cone variables. We may work with a more 
general metric of the form:
\eqn{genrlPen}{\eqalign{
ds^2 &= 2 du \,dv + \( f(u) \, x^2  
- \l^2 \, \vzs \) du^2 + dx^2
+ d\vzs + dy_5^2 \cr
\Phi(u) & = \phi(u); \qquad 2 {\partial^2 \phi(u) \over \partial u^2} = 
f(u) \cr 
B_{12} &= \l \, u
}}
String propagation in this geometry can be studied again in light-cone 
quantization with $u = \tau$. The sigma model is as discussed in the 
case of the extremal NS5-brane solution with the addition of the 
$f(u) x^2$ term arising from $g_{uu}$. 
If we write the sigma model in the light-cone, this leads to a 
time dependent mass term for the $x$ field. The five longitudinal 
directions of the five-brane are massless scalars and the coordinates 
along the $\R^2$ spanned by $(z^1, z^2)$ are free bosons modulo the 
shifted frequencies.

For the $x$-coordinate the world-sheet lagrangian reads:
\eqn{zlag}{
{1 \over 4 \pi \alpha'} \, \int d\tau \int_0^{2 \pi \alpha'
p^+} d\sigma \, \[ {1 \over 2} 
\left({\partial x \over \partial \tau}\right)^2 - {1 \over 2} 
\left({\partial x \over \partial \sigma} \right)^2 - { 1\over 2} f(\tau)\, 
x^2 \].
 }
The equation of motion for the $x$ field on the world-sheet, with
$x(\sigma, \tau) = e^{i n \sigma} x_n(\tau)$ and $u =  \tau$, is 
\eqn{xeom}{
\partial^2_\tau x_n(\tau) + n^2 x_n(\tau)  - f(\tau)\, x_n(\tau) = 0.
}
Although we have a non-trivial dilaton in the background, it does not 
explicitly enter the sigma model lagrangian, as the world-sheet curvature is 
zero in light-cone gauge. The dilaton affects the lagrangian by determining 
the function $f(\tau)$ in \zlag\ as in \genrlPen. In \genrlPen, the 
dilaton is determined up to a term linear in $u$, which has no physical 
effect on string propagation.

The above Lagrangian \zlag\ describes string propagation in a time dependent 
background, which was to be expected since we are looking for geodesics that 
are moving out from the horizon in \scalim. String propagation in such 
backgrounds was studied in \cite{horowitzs}. In such time-dependent 
backgrounds it was found that the string oscillators upon interacting with the 
plane wave can get excited. This is the benign form of particle production that
one can associate with these backgrounds. One can calculate the 
probabilities for finding the string at an excited state at some time, given 
its initial state at an earlier time.

We will interpret this time dependent background as {\it real time RG } 
in the little string theory. The motivation is that 
in the holographic description of the little string 
theory, the $\sigma$ direction in the background \scalim\ is the 
direction of the RG flow in the dual theory. This is the analog of the 
statement that radial evolution in AdS is equivalent to RG flow in the 
boundary conformal field theory \cite{dbdvv}.

The claim is then the following. The solution 
of the sigma model with given $u$ corresponds to 
studying the little string theory with a finite cut-off $\Lambda = u$. 
A simple consistency check is that in the limit of an infinite cut-off, 
$\Lambda = u \rightarrow \infty$, implying that the $x$ field is 
becoming massless on the world-sheet. This is precisely what we expect from
our study of high-energy of the little strings in the previous section. 

One might be worried about the singularity at $u =0$ in the metrics \Penrf\ 
and \Penrfl. This singularity actually corresponds to the black hole 
singularity in the original spacetime \scalim\ before taking the 
Penrose limit. One may see this explicitly by deriving the Penrose limit 
using the metric written in \nonextns. As mentioned earlier, the cigar 
geometry corresponds to little string theory at finite energy density 
$\mu_{lst}$. This means that there is an infra-red cut-off in the theory.
From the spacetime point of view, we should therefore restrict ourselves 
to the region outside the black hole horizon, as the radial position of 
the horizon determines the scale of the infra-red cut-off.
Thus, we should have $u \ge {1 \over \sqrt{1-\l^2}} \sinh^{-1}
\left({\sqrt{1-\l^2} \over \l} \right)$. It may be worrisome 
that such an ansatz leads to considering a geodesically incomplete spacetime. 
However, even if $u$ were allowed to go to $0$, the 
spacetime would still be geodesically incomplete, since then we would 
encounter a real singularity.

So what are the questions that can be answered in the little string theory 
from analysing string propagation on \Penrf\ or \Penrfl? 
We should quantize the theory \zlag\ as usual, by solving the equation of 
motion \xeom\ and imposing 
canonical commutation relations. One of the interesting questions we can 
ask is: Suppose we start in the vacuum state at $\Lambda = \infty$, 
what is the average occupation number of the string at any given cut-off?
One simple approach is to realize that the wave equation for the $x$-field 
\xeom\ can be mapped to a quantum mechanics problem in one-dimension wherein 
the energy is $n^2$ and the potential is $f(\tau)$. For small energies, 
{\it i.e.}, small $n$, compared with the height of the 
potential at the infra-red cut-off, use a WKB approximation to determine 
the average occupation number as
\eqn{avoccno}{
\langle N_{occ} \rangle = \exp \left( \int_{\Lambda}^{\Lambda_{UV}} \; 
\sqrt{f(\tau)}\,  d\tau \right) 
}
In fact, we can obtain an exact answer, without recourse to a 
WKB approximation, for the cases where $f(\tau)$ leads to a solvable 
quantum mechanical problem, such as $f(\tau) = \sech^2(\tau)$ and 
$f(\tau) = \tanh(\tau)$. The average occupation number is then determined 
in terms of the Bogoliubov coefficients \cite{horowitzs}.

For the metric \Penrf\ with $f(\tau) = {2 \over \tau^2}$ we can 
evaluate \avoccno\ to give  $\langle N_{occ} \rangle = 
\left({\Lambda_{UV} \over \Lambda}\right)^{\sqrt{2}}$. 
If we heuristically identify the average occupation 
number with the number of degrees of freedom of little string theory at the 
scale $\lambda$, we then conclude that there are more degrees of 
freedom in little string theory in the infra-red than in the ultra-violet.
While this would be bizarre for a local field theory, there are 
indications that such a feature might indeed hold for non-local theories.
 
In fact, for little string theory it is known from analysis of string 
propagation in the cigar geometry \cite{dvv}, that in addition 
to defining operators which are inserted at the asymptotic linear dilaton end, 
we can also have localized states near the tip of the cigar. 
These bound states show up as poles in correlation function of 
the operators \cite{gka,gkb}. 
If they are to be thought of as living close to the tip of the cigar, then 
we will not be able to see them in the ultra-violet, but by moving into the 
cigar we will begin to encounter these states.  But, this is precisely how we 
probe the geometry with the pp-wave, and therefore it shouldn't be 
surprising that we have more states as we reach the infra-red. 

\newsection{Interpolating between Little Strings and (2,0) SCFT}

In the previous section we considered a situation wherein we regulated the 
strong coupling region of the linear dilaton geometry by studying the 
two-dimensional black hole. However, our geodesic could still enter into 
the region behind the horizon, and the metric as written in \Penrf\ and 
\Penrfl\ had a region of strong coupling (not to mention a singularity)
in the region of small $u$. Our next aim is to better this and find a smooth
metric. We will take into account the effects of the strong 
coupling by studying the appropriate configuration in M-theory for the case 
of IIA little string theory.

\newsubsection{M5-brane array}

Type IIA NS5-branes have their origins in M-theory as M5-branes 
transverse to the M-theory circle. When we are looking at the 
near-horizon geometry of the Type IIA NS5-branes, we account 
for the strong coupling region of the resulting linear dilaton geometry by 
lifting the IIA solution to M-theory. So we need to write down the 
supergravity solution in 11 dimensions corresponding to an array of 
$N$ M5-branes localized at a point on the 11$^{\rm th}$ circle. 

The metric for this configuration is readily written down. The space 
transverse to the M5-branes is now $\R^4 \times \S^1$; we let $w$ 
be the radial coordinate on the $\R^4$ and $z_{11}$ 
be the direction along the M-theory circle. The metric then reads
\eqn{mfivemet}{
ds^2 = l_p^2 \( \tH^{-1/3} \, \[- dt^2 + dy_5^2 \] +  \tH^{2/3} \, \[
dz_{11}^2 + dw^2 + w^2 (\cos^2\theta \, d\psi^2 + d\theta^2 + 
\sin^2\theta \, d\phi^2) \] \) 
}
where
\eqn{Hdef}{ 
\tH = \sum_{n=-\infty}^{\infty} 
{N \over \[  w^2 + (z_{11} - {n \over l_s^2})^2 \]^{3/2}}}
We have already taken a near-horizon limit by dropping the additive 
constant in the definition of the harmonic function $\tH$.

The geometry described by the above metric is such that asymptotically we 
have a linear dilaton region, where the size of the 11$^{\rm th}$ circle is 
small and we can trust the 10-dimensional supergravity or perturbative 
string theory. As we come into the strong coupling region, the 11$^{\rm th}$
circle opens out and there is an intermediate region where there is 
a rather complicated geometry. As we further descend towards the location of 
the M5-branes we find a major simplification; the geometry is now simply the 
near-horizon geometry of a stack of M5-branes, which is simply $AdS_7 \times
\S^4$.

To see this explicitly, let us introduce new coordinates $\rho$ and $\chi$ 
where
\eqn{wzrhochi}{\eqalign{
w & = 4 N \rho^2 \sin(\chi) \cr
z_{11} & = 4 N \rho^2 \cos(\chi) 
}}
Under this change of coordinates the metric becomes
\eqn{mfnew}{\eqalign{
ds^2 & = 4 N^{2/3} l_p^2 \Bigg[ H^{-1/3} \( - dt^2 + dy_5^2 \) + \cr
 & \;\;\;\;\;\; + H^{2/3} \, \left\{
\rho^2 d\rho^2 + {\rho^4 \over 4} \( d\chi^2 +\sin^2\chi
\( \cos^2\theta \, d\psi^2 + d\theta^2 + 
\sin^2\theta \, d\phi^2 \) \) \right\} \Bigg] \cr 
H & = 4^3 \, N^2 \, \tH
= \sum_{n=-\infty}^{\infty} \; { 1 \over \[\rho^4 - 
{2 n\over \tilde{l}_s^2} 
\rho^2 \cos\chi + {n^2 \over \tilde{l}_s^4} \]^{3/2}}.
}}
with a rescaling $\tilde{l}_s = 2 \sqrt{N} l_s$. 
This change of coordinates, while not completely essential to deciphering the 
limiting nature of the metric \mfivemet, helps in simplifying the 
analysis and in  considering the Penrose limits of this geometry.

In the metric \mfivemet\ the Anti-de Sitter region arises when distances are 
small compared to the string length $l_s$. This means that we can 
simplify the harmonic function by collapsing the infinite sum to just the 
contribution from the $n=0$ piece. Then in the new coordinates \wzrhochi\ 
we find $H(\rho) = {1 \over \rho^6}$.  Direct substitution of $H$ into 
\mfnew\ yields the metric of $AdS_7 \times \S^4$,
\eqn{adsseven}{
ds^2 = 4 N^{2/3} l_p^2 \,\left[\rho^2 \left(-dt^2 + dy_5^2\right) + 
{d\rho^2 \over \rho^2} + {1 \over 4} d\Omega_4^2 \right]
}
The $\S^4$ is parameterized by $\chi$ and by the coordinates of the 
$\S^3$ appearing in \mfivemet.

In the opposite limit,
to extract the linear dilaton region of the array solution \mfivemet, we 
need to consider radial distances which are larger than all other 
distance scales, $w \gg l_p$.
As this limit is easier to visualize in the original $(w,z_{11})$ 
coordinates, let us first consider what happens to the limiting 
metric in these  before passing to the $(\rho,\chi)$ 
coordinates.

In the original geometry \mfivemet, we want to look at the region far 
away from the M5-branes, which means taking $w$ large.  From far away, 
the branes look uniformly smeared over the 11$^{ \rm th}$ direction; 
in other words the sum in \Hdef\ becomes an integral, {\it i.e.}, 
$\tH = {2N \over w^2}$.
The metric \mfivemet\ is then given by
\eqn{einsmetwz}{
ds^2 = l_p^2 (2N)^{-1/3} \, w^{2/3} \, \( - dt^2 +  dy_5^2 
+ {2N \over w^2} \, \[ dz_{11}^2 + dw^2 \] + 2N \, d\Omega_3^2 \).
}
We see that as $w \to \infty$, the $dz_{11}^2$ term drops out, 
and after further rescaling, the final metric is conformally related to 
the NS5-brane metric \nsmetric.

In the new $(\rho,\chi)$ coordinates, the same limit means taking
$\rho$ large, but also $\tan \chi \gg 1$, so that $\chi \sim {\pi \over 2}$.
In this limit, $H \sim {2 \over \rho^4}$, so that the metric 
becomes
\eqn{mfeinsmet}{
ds^2 = 2 l_p^2 (2N)^{2/3} \, \rho^{4/3} \, \( - dt^2 +  dy_5^2 
+ 2 \, {d \rho^2 \over \rho^2} + {1\over 2} \, \[ d\chi^2 + \sin^2 \chi \,
 d\Omega_3^2 \] \)
}
For $\chi \sim {\pi \over 2}$, the $\S^4$ part (in the square brackets) 
reduces to $d\Omega_3^2$, and the whole metric scales to the NS5-brane 
metric, as before. A point of some importance is that we naturally 
obtain the metric in the Einstein frame.

\newsubsection{Penrose limits and M5-array solution}

Our next step is to write down a pp-wave geometry which interpolates 
between the UV and the IR regions of the Type IIA little string theory. 
For this we will make use of the supergravity solution of an array of 
M5-branes in the form written in \mfnew. The essential ingredient is the 
identification of the right null geodesic. This is a null geodesic 
which is in the $(t, \rho, \psi)$ plane and is situated at 
$\chi = {\pi \over 2}$, $\theta = 0$, and $y_5 = 0$.
The tangent vector to the geodesic is given by
$\dot{t} \, \p_t  + \dot{\rho} \, \p_\rho +\dot{\psi} \, \p_\psi $. 
The existence of such a geodesic, despite the complicated dependence of 
the metric on the coordinate $\chi$ can be understood from the symmetry of 
the solution around $\chi = {\pi \over 2}$.

Using the constants of motion,
$E = H^{-1/3} \, \dot{t}$ and 
$J =  H^{2/3} \,{\rho^4 \over 4}  \, \dot{\psi}$ 
at $\chi={\pi \over 2}$ and $\theta = 0$, writing
$\l = J/E$ and rescaling $E = 1$, the geodesic equation, which may be 
obtained from \gengeod, reads
\eqn{mfgd}{
\rho^2 \dot{\rho}^2 =  H^{-1/3} \, \(1- {4 \l^2 \over \rho^4 H} \) }

Adapting coordinates to this null geodesic following \genctd\ 
then gives
\eqn{mfct}{\eqalign{
\rho \, d\rho &=   H^{-1/6} \, \(1- {4 \l^2 \over \rho^4 H} \)^{1/2} \, du \cr
d\psi &= {4 \l \over \rho^4  H^{2/3}} \, du + dx \cr
dt &= H^{1/3} \, du -dv + \l \, dx
}}
We can take the Penrose limit as discussed in Section 2, to obtain 
 the pp-wave metric in the Brinkman coordinates,
\eqn{genppM}{\eqalign{
ds^2 = & 2 \, du \, dv + dx^2 + d\vzs +
d\chi^2 + dy_5^2 + \cr
 & + \( - {4\l^2 \over \rho^4 H^{2/3}} \, \( \vzs + \chi^2 \) + 
{1 \over 2} f_{\vec{z}}(u) \, \vzs  +
{1 \over 2} f_{\chi}(u) \, \chi^2  +
{1 \over 2} f_{x}(u) \, x^2  + 
{1 \over 2} f_{5}(u) \, y_{5}^2   
\) \, du^2 
}}
where
\eqn{masstermsmf}{\eqalign{
f_i(u) &= {1 \over 2} \({\partial_u h_i(u) \over h_i(u)} \)^2 + 
\partial_u \( {\partial_u h_i(u) \over h_i(u)}\) \cr
h_5(u) & = H^{-1/3}(u) \cr
h_x(u) & = \( H^{2/3} {\rho^4 \over 4} - \l^2 H^{-1/3} \) \cr
h_{\vec{z}}(u) & = h_{\chi}(u) = { \rho^4 \over 4} H^{2/3}
}}
where we have written $H(u)$ as an implicit function of $u$. This may be 
determined by solving \mfgd\ for $\rho(u)$. 
We now demonstrate how to recover the limiting behaviour of the pp-wave 
metric \genppM. We will see that from this general form, we can isolate the 
pp-wave limit of the AdS region, as well as the linear dilaton region.

\newsubsection{Relating to the AdS pp-wave}

First we will recover the 11-dimensional pp-wave associated with the 
AdS region. Recall that to obtain the AdS region from the metric 
\mfnew\ we are to look at length scales small compared to the string 
scale. Therefore we have  $H(\rho) = {1 \over \rho^6}$. Plugging this 
into the metric in \mfnew, we recover the metric of $AdS_7 \times \S^4$ in 
Poincare coordinates as discussed earlier. 

Given $H(\rho) = {1 \over \rho^6}$, all we need to do is to solve for 
$\rho(u)$ using \mfgd. We have
\eqn{adsgeo}{
\rho \, d\rho = \rho \(1 - 4 \l^2 \rho^2\) du \Rightarrow \rho(u) = 
{1 \over 2\l}
\sin(2 \l u)
}
This implies that the functions $h_i(u)$, which determine the 
explicit form of the pp-wave metric as in \masstermsmf, are given as 
\eqn{adsh}{\eqalign{
h_5(u) &= \rho^2(u) = { 1 \over 4 \l^2} \sin^2( 2 \l u) \cr
h_x(u) &= {1 \over 4} \cos^2(2 \l u) \cr
h_{\vec{z}}(u) &= h_{\chi}(u) = 
{1 \over 4}.
}}
Therefore the resulting metric is:
\eqn{adslmet}{
ds^2 =  2\, du \, dv -  \l^2 \, \(\, 4 \, y_5^2 + 4\, x^2 + \vzs +
\chi^2 \) \, du^2 + dy_5^2 + dx^2 + d\vzs + d\chi^2
}
which we recognize as the usual pp-wave limit of $AdS_7 \times \S^4$, 
{\it cf.}, \cite{Bmn}.

\newsubsection{Relating to the linear dilaton pp-wave}

Let us now turn to the other limit of interest, 
namely the linear dilaton regime. This we recall is obtained by
looking at the array of M5-branes from afar {\it i.e.}, $\rho \gg 1$.
Here we had $ H \sim {2 \over \rho^4}$, so we can again solve
for $\rho(u)$ using \mfgd, obtaining
\eqn{ldru}{
\rho \, d\rho = {\rho^{2/3} \over 2^{1/6}} \( 1 - 2 \l^2 \)^{1/2} \, du. 
}
Hence $\rho = \( {4 \over 3} 2^{-1/6} \, \sqrt{1 - 2 \l^2} \, u \)^{3/4}$, 
so that $H={a^3 \over u^3}$ where $a = { 3 \over 2^{3/2} \sqrt{1 - 2 \l^2}}$, 
and $ {4 \l^2 \over \rho^4 H^{2/3}} = {b \over u}$ with 
$b = { 3 \, \l^2 \over 2^{1/2} \, \sqrt{1 - 2 \l^2}}$.

For the functions $h_i(u)$ which are crucial in determining the pp-wave 
metric as in \masstermsmf, we have:
\eqn{ldhf}{\eqalign{
h_5(u) = {u \over a} \ \ \ & \Rightarrow \ \ \ f_5(u) = - {1 \over 2 u^2} \cr
h_x(u) = u \, \l^2 \, \({1 \over b} - {1 \over a}\)
 \ \ \ & \Rightarrow \ \ \ f_x(u) = - {1 \over 2 u^2} \cr
h_{\vec{z}}(u) = h_{\chi}(u) = {u \, \l^2 \over b} \ \ \ & \Rightarrow \ \ \ 
f_{\vec{z}}(u) = f_{\chi}(u) = - {1 \over 2 u^2}.
}}
With $\chi=0, d\chi =0$, this yields the metric
\eqn{eldppwave}{
ds^2 = 2 \, du \, dv + dx^2 + d\vzs + dy_5^2 -
{1 \over 4u^2} \(\vzs + x^2 + y_5^2 + 4 b \, u \, \vzs \) du^2}

At first sight this doesn't appear to resemble the formulae for the pp-wave 
metric obtained in Section 3. This is not much of a surprise, for the 
M-theory solution naturally descends into the Einstein frame metric 
for the Type IIA NS5-branes, which is related to the string frame 
metric appearing in \nsmetric\ through a conformal transformation. In fact, 
the Einstein frame metric is given in \einsmetwz\ or \mfeinsmet. 
Although conformal transformations do not change the geodesic equation, 
they do affect the Penrose limit, since we are making use of conserved 
quantities to parametrize our solutions to the geodesic equation. 
They are affected because conformal transformations involve a local 
recalibration of our measuring devices. If we write down the Penrose limit 
of the geometry \mfeinsmet, for a null geodesic which carries 
angular momentum along the $\S^3$ and also at the same time is falling in
into the linear dilaton region, we will
indeed obtain the pp-wave metric \eldppwave.

This demonstrates that the pp-wave geometry given in \masstermsmf\ is a 
plane wave geometry interpolating between the linear dilaton and the 
AdS plane waves, providing an explicit supergravity solution which can be 
used to study the properties of renormalization group flow in little string 
theory in the absence of an infra-red cut-off. Notice that unlike the 
case of the cigar geometry, in the infra-red region,
there is now no singularity in the solution.

\newsection{Non-commutative theories}

Non-commutative field/string theories provide yet another example of 
non-local theories; one might naturally wonder whether considerations of 
Penrose limits of the 
corresponding holographic dual spacetime leads to some useful 
statements about these theories. We will try to study RG flows in these 
theories too, by considering geodesics with a radial component. 
Unfortunately, the resulting pp-wave geometry will prove to be 
quite messy and does not lend any insight into the physics of 
non-commutative theories. 

As our final example, we therefore consider the Penrose limit of the
supergravity dual to non-commutative Yang-Mills theories in four dimensions. 
We shall look at the near-horizon geometry of D3-branes with a constant 
NS-NS B-field on their world-volume. The supergravity solution 
for this geometry can be found for instance in \cite{hait} or in \cite{maru}, 
{\it c.f.}, Eq (2.7) of the latter.
We write the metric as:
\eqn{ncmet}{
ds^2 = R^2 \[ -r^2 \, dt^2 + { dr^2 \over r^2} + \cos^2\theta \, d\psi^2
+ d\theta^2 + \sin^2\theta \, d\Omega_3^2 + r^2 \, dy_1^2 +
{ r^2 \over 1 + a^4 r^4 } \,  dy_2^2 \]
}
Note that this differs from the conventions of \cite{maru} for we have 
replaced their coordinates $u \rightarrow r$,
$\tilde{x}_1 \rightarrow y_1$, $(d\tilde{x}_2^2 + d\tilde{x}_3^2 ) \rightarrow
dy_2^2$, in addition to setting $\alpha' =1$. In this case,
$R^2 \sim \sqrt{N}$ is taken to be large.

We might wish to take a Penrose limit along a null geodesic in the 
$(t, \psi)$ plane, staying at constant $r$, in analogy with the 
considerations in \cite{Bmn}. However, we will soon see that such 
a geodesic does not exist in this case.
Therefore, let us consider the more general geodesic in the
$(t, r, \psi)$ plane, which moves both ``radially'' and along the equator of
the $\S^5$, at $\theta = 0$, $x_i = 0$.
The tangent vector to the geodesic is 
$\dot{t} \, \p_t + \dot{r} \, \p_r  +\dot{\psi} \, \p_\psi$, 
and gives the constants of 
motion $E \equiv r^2 \, \dot{t}$ and $J \equiv \dot{\psi}$ at $\theta = 0$.
We obtain the geodesic equation as in \gengeod
\eqn{ncge}{
\dot{r}^2 = 1- \l^2 \, r^2
}
Note that this is exactly the same geodesic equation as one would find 
for $AdS_5 \times \S^5$ in the Poincare coordinates, as it should be.

From the effective potential, it is now obvious that the geodesic can't 
stay at any constant $r>0$.
In fact, one can easily integrate the geodesic equation to find
\eqn{ncgeod}{
r(u) =  {1 \over \l} \sin(\l \, u)
}
which in the limit $\l \to 0$ correctly reduces to $r(u) =  u$.

Let us therefore consider this more general geodesic, and
construct the corresponding pp-wave metric. The change of coordinates 
suggested by \genctd\ is
\eqn{ncct}{\eqalign{
r &= {1 \over \l} \sin(\l \, u)  \cr
t &= - \l \, \cot(\l \, u) - v + \l \, x \cr
\psi &= \l \, u + x
}}
which, after substitution and rescaling as in \coordres\ and further 
coordinate transform to cast the resulting pp-wave metric in Brinkman 
form, leads to the metric
\eqn{ncpp}{\eqalign{
ds^2 = 2 \, du \, dv 
&- \l^2 \, \(x^2 + \vzs + y_1^2 + g(u) \, y_2^2\) \, du^2 + \cr
&+ dx^2 + d\vzs + dy_1^2 + dy_2^2,
}}
where
\eqn{ncppg}{
 g(u) \equiv {\l^8 - a^8 \sin^8(\l u) - 2 a^4 \sin^2(\l u) \, \cos^2(\l u)
\( a^4 \sin^4(\l u) - 5 \l^4 \) \over
\( \l^4 +  a^4 \sin^4(\l u) \)^2 .}}

Several further comments are in order. First, setting $a=0$, we recover the 
maximally supersymmetric pp-wave resulting from $AdS_5 \times \S^5$.
Also we recover the same for small $u$, for in this limit, 
the distinction between the $y_1$ and the $y_2$ directions
disappears. In both these cases, we are considering essentially the 
analog of the vanishing B-field. We will not write down the 
various form fields and the dilaton which are non-vanishing 
in the supergravity background \ncmet\ and simply note that they lead to 
non-vanishing and finite fields in the pp-wave background as guaranteed by 
the general properties of the Penrose limit \cite{gueven}.

Unfortunately, string theory in the background \ncppg\ remains 
intractable even in the light-cone gauge. This seriously hampers 
our initial objective of hoping to understand the features of RG flow in 
the non-commutative field theory using the Penrose limit. 

\newsection{Conclusion}

In this paper we have studied various Penrose limits of string theory 
backgrounds that are dual to little string theory and non-commutative gauge 
theory. The resulting pp-wave metrics are, by construction, solutions to the 
low energy effective action of the type II string theory, and thus provide 
consistent backgrounds for string propagation. By studying strings
in these backgrounds one obtains information about specific sectors of 
the corresponding non-local theory. 
In particular, for little string theory
we have identified a sector with high energy and $R$-charge that is dual to 
strings in the Nappi-Witten background, with a spectrum that is essentially 
that of a free string theory. It would of course be nice to verify this 
correspondence explicitly in a similar way as for $N=4$ super Yang-Mills, but
at present this is hard to do since very little is known about little string 
theory. 

For little strings with a finite energy density we derived a one-parameter 
family of Penrose limits of the dual `cigar' geometry. 
An interesting feature of these pp-waves is the explicit dependence on the 
light-cone coordinate. As a result, the world-sheet theory of strings 
in these backgrounds is described in the light-cone gauge by scalars and 
fermions with mass terms that explicitly depend on the world-sheet time.
{}From a dual perspective this time dependence is interpreted as a 
time-dependent RG-scale. String propagation in these
backgrounds can be studied qualitatively using  semi-classical methods.
It would be interesting to perform a more quantitative analysis of
these backgrounds and illuminate the dual RG-interpretation in more
detail.

Another example, with an obvious dual RG-interpretation, is the 
pp-wave geometry that interpolates between the `static' pp-waves 
corresponding to little strings in the UV and the (2,0) theory in the IR. 
The somewhat complicated form of this pp-wave
metric, as well as the one corresponding to non-commutative gauge theory, makes
a detailed analysis of the string theory very hard. We expect that
the methods discussed in this 
paper, when applied to other backgrounds
with dual interpretations, will yield pp-wave metrics with similar 
time-dependence. For example, one could consider the D1-D5 system and 
other intersecting brane configurations. 
It would be very interesting if
among these one
can find examples that leads to  `solvable' (=integrable) time-dependent
world-sheet theories.

\section*{Acknowledgments}

We would like to thank Sameer Murthy and  Nati Seiberg for interesting 
discussions. The work of VH and MR is supported by NSF grants PHY-9870115 and 
PHY-9802484 respectively, while EV is supported by a DOE grant
DE-FG02-91ER40571.



\begin{thebibliography}{99}
\bibitem{cds}
A.~Connes, M.~R.~Douglas and A.~Schwarz,
``Noncommutative geometry and matrix theory: Compactification on tori,''
JHEP {\bf 9802}, 003 (1998)
[arXiv:hep-th/9711162].

\bibitem{dh}
M.~R.~Douglas and C.~M.~Hull,
``D-branes and the noncommutative torus,''
JHEP {\bf 9802}, 008 (1998)
[arXiv:hep-th/9711165].

\bibitem{sw}
N.~Seiberg and E.~Witten,
``String theory and noncommutative geometry,''
JHEP {\bf 9909}, 032 (1999)
[arXiv:hep-th/9908142].

\bibitem{sst}
N.~Seiberg, L.~Susskind and N.~Toumbas,
``Strings in background electric field, space/time noncommutativity  
and a new noncritical string theory,''
JHEP {\bf 0006}, 021 (2000)
[arXiv:hep-th/0005040].

\bibitem{gmms}
R.~Gopakumar, J.~M.~Maldacena, S.~Minwalla and A.~Strominger,
``S-duality and noncommutative gauge theory,''
JHEP {\bf 0006}, 036 (2000)
[arXiv:hep-th/0005048].

\bibitem{brs}
M.~Berkooz, M.~Rozali and N.~Seiberg,
``On transverse fivebranes in M(atrix) theory on T**5,''
Phys.\ Lett.\ B {\bf 408}, 105 (1997)
[arXiv:hep-th/9704089].

\bibitem{seiberg}
N.~Seiberg,
``New theories in six dimensions and matrix 
description of M-theory on  T**5 and T**5/Z(2),''
Phys.\ Lett.\ B {\bf 408}, 98 (1997)
[arXiv:hep-th/9705221].

\bibitem{abks}
O.~Aharony, M.~Berkooz, D.~Kutasov and N.~Seiberg,
``Linear dilatons, NS5-branes and holography,''
JHEP {\bf 9810}, 004 (1998)
[arXiv:hep-th/9808149].

\bibitem{hait}
A.~Hashimoto and N.~Itzhaki,
``Non-commutative Yang-Mills and the AdS/CFT correspondence,''
Phys.\ Lett.\ B {\bf 465}, 142 (1999)
[arXiv:hep-th/9907166].

\bibitem{maru}
J.~M.~Maldacena and J.~G.~Russo,
``Large N limit of non-commutative gauge theories,''
JHEP {\bf 9909}, 025 (1999)
[arXiv:hep-th/9908134].

\bibitem{minsei}
S.~Minwalla and N.~Seiberg,
``Comments on the IIA NS5-brane,''
JHEP {\bf 9906}, 007 (1999)
[arXiv:hep-th/9904142].

\bibitem{NM}
K.~Narayan and M.~Rangamani,
``Hot little string correlators: A view from supergravity,''
JHEP {\bf 0108}, 054 (2001)
[arXiv:hep-th/0107111].

\bibitem{penrose}
R.~Penrose,
``Any spacetime has a planewave as a limit,'' in
{\it Differential geometry and relativity}, pp 271-275,
Reidel, Dordrecht, 1976.

\bibitem{metsaev}
R.~R.~Metsaev,
``Type IIB Green-Schwarz superstring in plane wave Ramond-Ramond  background,''
Nucl.\ Phys.\ B {\bf 625}, 70 (2002)
[arXiv:hep-th/0112044].

\bibitem{Bmn}
D.~Berenstein, J.~Maldacena and H.~Nastase,
``Strings in flat space and pp waves from N = 4 super Yang Mills,''
[arXiv:hep-th/0202021].

\bibitem{tseytlin}
A.~A.~Tseytlin,
``On limits of superstring in AdS(5) x S**5,''
[arXiv:hep-th/0201112].

\bibitem{gueven}
R.~Gueven,
``Plane wave limits and T-duality,''
Phys.\ Lett.\ B {\bf 482}, 255 (2000)
[arXiv:hep-th/0005061].

\bibitem{blaua}
M.~Blau, J.~Figueroa-O'Farrill, C.~Hull and G.~Papadopoulos,
``A new maximally supersymmetric background of IIB superstring theory,''
JHEP {\bf 0201}, 047 (2002)
[arXiv:hep-th/0110242].

\bibitem{blaub}
M.~Blau, J.~Figueroa-O'Farrill, C.~Hull and G.~Papadopoulos,
``Penrose limits and maximal supersymmetry,''
[arXiv:hep-th/0201081].

\bibitem{blauc}
M.~Blau, J.~Figueroa-O'Farrill and G.~Papadopoulos,
``Penrose limits, supergravity and brane dynamics,''
[arXiv:hep-th/0201081].

\bibitem{rosen}
N.~Rosen, Phys.\ Z.\ Sowjet.\ {\bf 12}, 366 (1937).

\bibitem{robinson}
I.~Robinson, Report to the Eddington Group, Cambridge (1956).

\bibitem{brinkman}
H.~W.~Brinkman, Proc.\ Natl.\ Acad.\ Sci.\ (U.S.) {\bf 9}, 1 (1923).

\bibitem{amkl}
D.~Amati and C.~Klimcik,
``Nonperturbative Computation Of The Weyl Anomaly For A 
Class Of Nontrivial Backgrounds,''
Phys.\ Lett.\ B {\bf 219}, 443 (1989).

\bibitem{horst}
G.~T.~Horowitz and A.~R.~Steif,
``Space-Time Singularities In String Theory,''
Phys.\ Rev.\ Lett.\  {\bf 64}, 260 (1990).

\bibitem{nappiw}
C.~R.~Nappi and E.~Witten,
``A WZW model based on a nonsemisimple group,''
Phys.\ Rev.\ Lett.\  {\bf 71}, 3751 (1993)
[arXiv:hep-th/9310112].

\bibitem{googuri}
J.~Gomis and H.~Ooguri,
``Penrose limit of N = 1 gauge theories,''
[arXiv:hep-th/0202157].

\bibitem{kp}
E.~Kiritsis and B.~Pioline,
``Strings in homogeneous gravitational waves and null holography,''
arXiv:hep-th/0204004.

\bibitem{kk}
E.~Kiritsis and C.~Kounnas,
``String Propagation In Gravitational Wave Backgrounds,''
Phys.\ Lett.\ B {\bf 320}, 264 (1994)
[Addendum-ibid.\ B {\bf 325}, 536 (1994)]
[arXiv:hep-th/9310202].


\bibitem{kkl}
E.~Kiritsis, C.~Kounnas and D.~Lust,
``Superstring gravitational wave backgrounds with space-time supersymmetry,''
Phys.\ Lett.\ B {\bf 331}, 321 (1994)
[arXiv:hep-th/9404114].


\bibitem{fhhp}
P.~Forgacs, P.~A.~Horvathy, Z.~Horvath and L.~Palla,
``The Nappi-Witten string in the light-cone gauge,''
Heavy Ion Phys.\  {\bf 1}, 65 (1995)
[arXiv:hep-th/9503222].


\bibitem{russota}
J.~G.~Russo and A.~A.~Tseytlin,
``Heterotic strings in uniform magnetic field,''
Nucl.\ Phys.\ B {\bf 454}, 164 (1995)
[arXiv:hep-th/9506071].

\bibitem{russotb}
J.~G.~Russo and A.~A.~Tseytlin,
``On solvable models of type IIB superstring in NS-NS and R-R plane wave  
backgrounds,''
JHEP {\bf 0204}, 021 (2002)
[arXiv:hep-th/0202179].

\bibitem{horowitzs}
G.~T.~Horowitz and A.~R.~Steif,
``Strings In Strong Gravitational Fields,''
Phys.\ Rev.\ D {\bf 42}, 1950 (1990).

\bibitem{malstr}
J.~M.~Maldacena and A.~Strominger,
``Semiclassical decay of near-extremal fivebranes,''
JHEP {\bf 9712}, 008 (1997)
[arXiv:hep-th/9710014].

\bibitem{zayas}
L.~A.~P.~Zayas and J.~Sonnenschein,
``On Penrose limits and gauge theories,''
JHEP {\bf 05}, 010 (2002)
[arXiv:hep-th/0202186].

\bibitem{dbdvv}
J.~de Boer, E.~Verlinde and H.~Verlinde,
``On the holographic renormalization group,''
JHEP {\bf 0008}, 003 (2000)
[arXiv:hep-th/9912012].

\bibitem{dvv}
R.~Dijkgraaf, H.~Verlinde and E.~Verlinde,
Nucl.\ Phys.\ B {\bf 371}, 269 (1992).

\bibitem{gka}
A.~Giveon and D.~Kutasov,
``Little string theory in a double scaling limit,''
JHEP {\bf 9910}, 034 (1999)
[arXiv:hep-th/9909110].

\bibitem{gkb}
A.~Giveon and D.~Kutasov,
``Comments on double scaled little string theory,''
JHEP {\bf 0001}, 023 (2000)
[arXiv:hep-th/9911039].



\end{thebibliography}
\end{document}